\documentclass{article}

\usepackage{microtype}
\usepackage{graphicx}
\usepackage{subfigure}
\usepackage{booktabs} 

\usepackage{hyperref}

\usepackage[accepted]{mlsys2020}

\usepackage{amsmath,amssymb,amsfonts}
\usepackage{textcomp}
\usepackage{xcolor}
\usepackage{fancyhdr}
\usepackage{import}
\usepackage{mathtools}
\usepackage{comment}

\DeclarePairedDelimiter\ceil{\lceil}{\rceil}

\mlsystitlerunning{Rethinking Floating Point Overhead}
\begin{document}
\twocolumn[
\mlsystitle{Rethinking Floating Point Overheads\\ for Mixed Precision DNN Accelerators}


\mlsyssetsymbol{equal}{*}
\begin{mlsysauthorlist}
\mlsysauthor{Hamzah Abdel-Aziz}{ssi}
\mlsysauthor{Ali Shafiee}{ssi}
\mlsysauthor{Jong Hoon Shin}{ssi}
\mlsysauthor{Ardavan Pedram}{ssi}
\mlsysauthor{Joseph H. Hassoun}{ssi}
\end{mlsysauthorlist}

\mlsysaffiliation{ssi}{Samsung Semiconductor, Inc. San Jose, CA}

\mlsyscorrespondingauthor{Hamzah Abdel-Aziz}{hamzah.a@samsung.com}

\mlsyskeywords{Machine Learning, Hardware Acceleration}

\vskip 0.3in

\begin{abstract}

In this paper, we propose a mixed-precision convolution unit architecture which supports different integer and floating point~(FP) precisions.
The proposed architecture is based on low-bit inner product units and realizes higher precision based on temporal decomposition. 
We illustrate how to integrate FP computations on integer-based architecture and evaluate overheads incurred by FP arithmetic support.
We argue that alignment and addition overhead for FP inner product can be significant since the maximum exponent difference could be up to 58 bits, which results into a large alignment logic.
To address this issue, we illustrate empirically that no more than 26-bit product bits are required and up to 8-bit of alignment is sufficient in most inference cases.
We present novel optimizations based on the above observations to reduce the FP arithmetic hardware overheads.
Our empirical results, based on simulation and hardware implementation, show significant reduction in FP16 overhead.
Over typical mixed precision implementation, the proposed architecture achieves area improvements of up to 25\% in TFLOPS/$mm^2$ and up to 46\% in TOPS/$mm^2$ with power efficiency improvements of up to 40\% in TFLOPS/W and up to 63\% in TOPS/W.
\end{abstract}
]




\printAffiliationsAndNotice{}  


\section{Introduction}\label{S:intro}

Deep Neural Networks (DNNs) have shown tremendous success in modern AI tasks such as computer vision, natural language processing, and recommender systems~\cite{lecun2015deep}. 
Unfortunately, DNNs success comes at the cost of significant computational complexity (e.g., energy, execution time etc.).
Therefore, DNNs are accelerated on specialized hardware units (DNN accelerators) to improve both performance and energy efficiency~\cite{jouppi2017datacenter,tensorcore,reuther2019survey}. 
DNN accelerators may utilize quantization schemes to reduce DNNs memory footprint and computation time~\cite{deng2020model}.
A typical quantization scheme compresses all DNN's layers into the same low-bit integer, which can be sub-optimal, as different layers have different redundancy and feature distributions~\cite{Wang_2019_CVPR,wu2018mixed}.
On the other hand, mixed precision quantization scheme assigns different precisions (i.e., bit width) for different layers and it shows remarkable improvement over uniform quantization~\cite{song2020drq,Wang_2019_CVPR,chu2019mixed,cai2020zeroq}.
Therefore, mixed-precision quantization schemes~\cite{song2020drq,Wang_2019_CVPR,chu2019mixed,cai2020zeroq} or hybrid approaches where a few layers are kept in FP and the rest are quantized to integer are considered to maintain FP32-level accuracy~\cite{zhu16,venkatesh2017accelerating}.

Half precision floating point (FP16) and custom floating point data types (e.g., bfloat16~\cite{abadi2016tensorflow}) are adopted for inference and training in several cases when quantization is not feasible (online learning, private dataset, supporting legacy code ... etc.).
They could reduce memory footprint and computation by a factor of two, without significant loss of accuracy and they are often obtained by just downcasting the tensors. 
FP16 shows remarkable benefits in numerous DNN training applications where FP16 is typically used as the weights and activation data type and FP32 is used for accumulation and gradient update~\cite{micikevicius2017mixed,jia2018highly, ott2019fairseq}. 

Data precision varies significantly from low-bit integer to FP data types (e.g., INT4, INT8, FP16, etc.) within or across different DNN applications.
Therefore, mixed-precision DNN accelerators that support versatility in data types are crucial and sometimes mandatory to exploit the benefit of different software optimizations (e.g., low-bit quantization).
Moreover, supporting versatility in data types can be leveraged to trade off accuracy for efficiency based on the available resources~\cite{shen2020fractional}.
Typically, mixed-precision accelerators are designed based on low precision arithmetic units, and higher precision operation can be supported by fusing the low precision arithmetic units temporally or spatially.


The computation of DNNs boils down to the dot product as the basic operation.
Typically, inner product is implemented either by temporally exploiting a multiply-accumulate (MAC) unit in time or in space using an inner product (IP) unit with multipliers followed by an adder tree. 
The multiplier and adder bit widths are the main architectural decisions in implementing the arithmetic unit to implement the dot product operation.
The multiplier precision is a key factor for the final performance, and efficiency for both IP and MAC based arithmetic units.
For example, a higher multiplier precision (e.g., $8\times8$) limits the benefit of lower-bit (e.g., INT4) quantization.
On the other hand, while lower precision multipliers are efficient for low-bit quantization, they incur excessive overhead for the addition units. 
Therefore, multipliers bit width is decided based on the common case quantization bit width.
The adder bit width in integer IP based architecture matches the multiplier output bit width. Thus, they can improve energy efficiency by using smaller adder and sharing the accumulation logic.
However, in multiply-and-accumulate (MAC) based architectures~\cite{chen2016eyeriss}, adders are larger to serve as accumulators as well. This overhead is more pronounced in low-power accelerators with low-precision multipliers optimized for low-bit quantized DNNs.

Implementing a floating point IP (FP-IP) operation requires alignment of the products before summation, which require large shift units and adders.
Theoretically, the maximum range of alignment between FP16 products requires shifting the products up to 58-bit.
Thus, the adder tree precision (i.e., bit width) to align any two FP16 products would impose an additional 58 bits in its input precision.
Such alignments are only needed for FP operations and appear as significant power and area overhead for INT operations, especially when IP units are based on low-precision multipliers.

In this paper, we explore the design space trade-offs of IP units that support both FP and INT based convolution.
We make a case for a dense low-power convolution unit that intrinsically supports INT4 operations.
Furthermore, we go over the inherent overheads to support larger INT and FP operations.
We consider INT4 for two main reasons. First, this data type is the smallest type supported in several modern architectures that are optimized for deep learning (e.g., AMD MI50~\cite{amdIi50}, Nvidia Turing architecture~\cite{kilgariff2018nvidia} and Intel Sprig Hill~\cite{springhill}). 
Second, recent research on quantization report promising results for 4-bit quantization schemes~\cite{ fang2020post, jung2019learning, nagel2020up, Choukroun19, banner2019post, Wang_2019_CVPR, choi2018pact, zhuang2020training}.
In spite of this, the proposed optimization is not limited to INT4 case and can be applied for other cases (e.g., INT8) as we discuss in Section~\ref{sec:method}.

The contributions of the paper are as follows:
\begin{enumerate}
    \item We investigate approximated versions of FP-IP operation with limited alignments capabilities. 
    We derive the mathematical bound on the absolute error and conduct numerical analysis based on DNN models and synthetic values. 
    We postulate that approximate FP-IP can maintain the GPU-based accuracy if it can align the products by at least 16 bits and 27 bits, for FP16 and FP32 accumulators, respectively.
    \item We demonstrate how to implement large alignments using smaller shift units and adders in multiple cycles. 
    This approach decouples software accuracy requirements from the underlying IP unit implementation. 
    It also enables more compact circuits at the cost of FP task performance.
    \item Instead of running many IP units synchronously in one tile, we decompose them into smaller clusters. 
    This can isolate FP-IP operations that need a large alignment and limits the performance degradation to one cluster.
    \item 
    We study the design trade-offs of our architecture. 
\end{enumerate}
The proposed architecture, implemented in standard 7nm technology, can achieve up to 25\% in TFLOPS/$mm^2$ and up to 46\% in TOPS/$mm^2$ in area efficiency and up to 40\% in TFLOPS/W and up to 63\% in TOPS/W in power efficiency. 

The rest of this paper is organized as follows. 
In Section~\ref{sec:ipu}, we present the proposed architecture of mixed-precision inner product unit (IPU) and explain how it can support different data types including FP16.
In section~\ref{sec:fp_op}, we first review the alignment requirement for FP16 operations and offer architecture optimization to reduce FP16 overheads.
Section~\ref{sec:method} goes over our methodology and discusses the empirical results.
In Section~\ref{sec:related}, we review related work, and we conclude the paper in Section~\ref{sec:conclusion}.

\section{Mixed-precision Inner Product Unit}
\label{sec:ipu}
To support different types of data types and precisions, we use a fine-grain convolution unit that can run INT4 intrinsically and realize larger sizes temporally.
We consider INT4 as the default common case since several recent research efforts are promoting INT4 quantization schemes for efficient inference~\cite{jung2019learning, nagel2020up}. However, the proposed architecture can be applied to other cases such as INT8 as the baseline.

\begin{figure}[!t]
\vspace{-10pt}
\begin{center}
   \includegraphics[width=.9\linewidth]{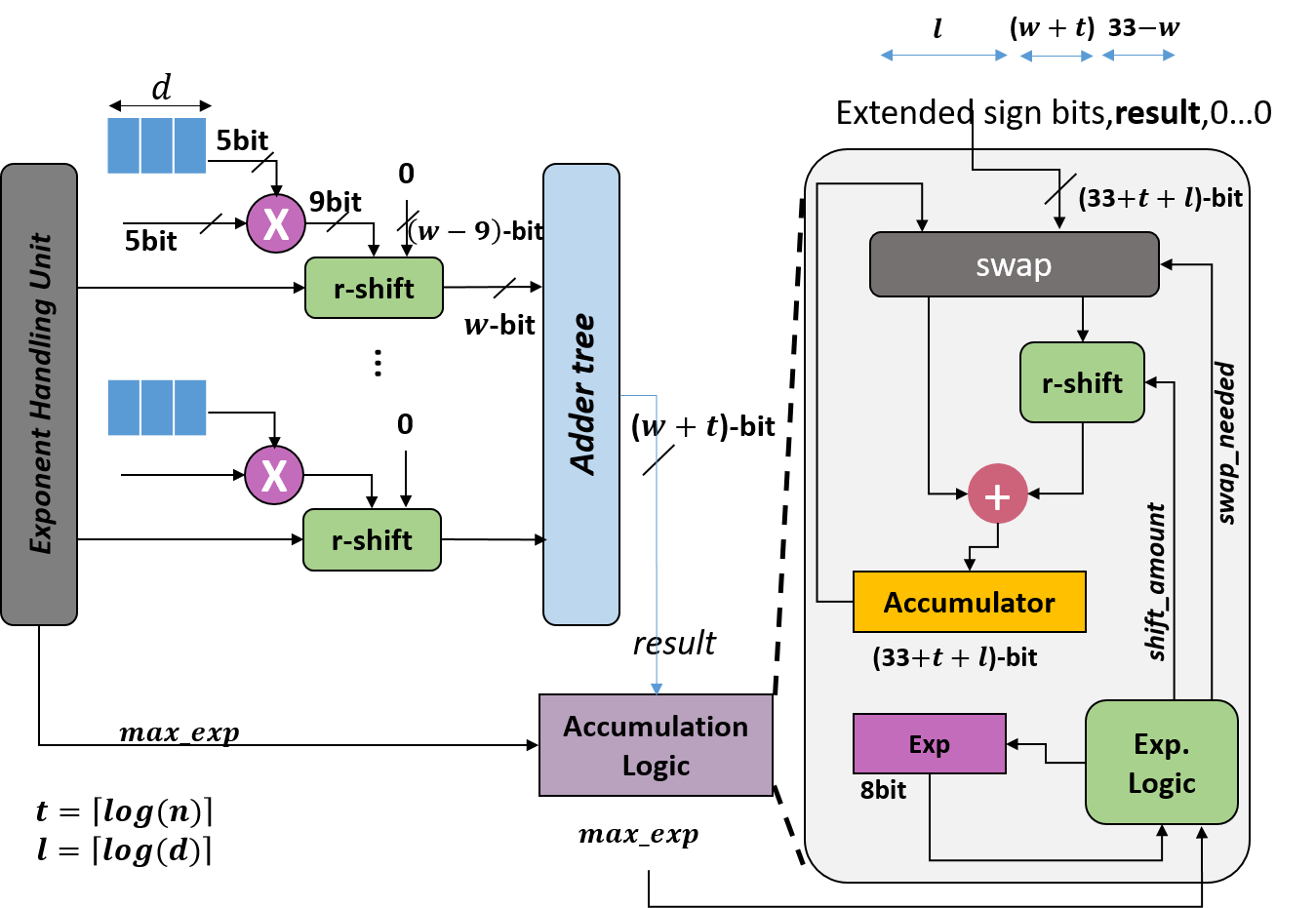}
\end{center}
   \vspace{-18pt}
   \caption{Microarchitecture of the proposed mixed-precision IPU data path with $n$ inputs and \emph{w}-bit IPU precision. }
\vspace{-15pt}
\label{fig:IPU}
\end{figure}

Figure~\ref{fig:IPU} shows the building blocks of the proposed mixed-precision $n$-input IPU, which is based on 5b$\times$5b sign multipliers. 
The proposed IPU allows computing INT4 IPU multiplications, both signed or unsigned, in a single cycle. 
In addition, larger precision operations can be computed in multiple \textbf{\emph{nibble iterations}}. 
The total number of nibble iterations is the multiplication of the number of nibbles of the two multipliers operands.  
Products are passed to a local right shift unit which used in FP-mode for alignment, and the shifted outputs are connected to an adder tree. The adder tree results are fed to the accumulator. 
In the next two subsection, we illustrate the mircoarchitecture in details for both INT and FP modes; respectively.

\subsection{INT Mode}
\label{sec:int-mode}
IPU is based on INT4 and the computation of higher INT precision is based on \textbf{\emph{nibble iterations}}.
For example, if the multipliers operands are INT8 and INT12, thus six nibble iterations are required to complete INT8 by INT12 multiplication for a single IP operation.
The local shift amount is always $0$ since there is no alignment required in INT mode. 
The result of the adder tree is concatenated with $(33-w)$ bits of zeros on the right side and always fed to the accumulator shift unit through the swape unit. The amount of shift depends on the significance of the nibble operands. 
For instance, suppose $N_k$ refers to the nibbles of a number (i.e., $N_0$ is the least significant nibble), the amount of shift for the result of IPU operation of nibble $N_i$ and $N_j$ for the first and the second operands is $4\times((K_a - i - 1)+(K_b - j - 1))$, where $K_a$ and $K_b$ are the total number of nibbles for operand $a$ and $b$, respectively.
The accumulator can add up to $n \times d$ multiplications, where $n$ is the number of IPU inputs and $d$ is the maximum number of times IPU can accommodate accumulation without overflow. In this scenario, the accumulator size should be at least $33+t+l$, where $l=\ceil*{\log_2 d}$.  
In INT mode, we assume $exp=max\_exponent=0$.

\subsection{FP Mode}
\label{sec:fp-mode}
In FP-mode, the mantissa multiplication is computed similar to INT12 IPU operation but with the following additional operation. 

\textbf{Converting numbers:} 
Let's define the magnitude of FP number as 0.mantissa for subnormal and 1.mantissa for normal FP numbers. We also call it the signed magnitude when sign bit are considered. Suppose $M[11:0]$ is the 12-bit signed magnitude for the FP16 number, it is converted to the following three 5-bit nibble operands: $N_2=\{M_{11}-M_7\}$, $N_1=\{0,M_6-M_3\}$, and $N_0=\{0,M_2-M_0,0\}$.
This decomposition introduces a zero in the least significant position of $N_0$. Since the FP-IP operation relies on right shifting and truncation to align the products, the implicit left shift of operands can preserve more accuracy.

\textbf{Local alignment:} The product results should be aligned with respect to the maximum exponent of all products (see Appendix~\ref{sec:background} for more details). 
Therefore, each of the multiplier outputs is passed to a local right shift unit that receives the shift amount from \textbf{\emph{the exponent handling unit}} (\textbf{EHU}). The EHU computes the product exponents by doing the following steps, in order: (1) element-wise summation of the operands' unbiased exponents, (2) computing the maximum of the product exponents, and (3) computing the alignment shift amounts as the difference between all the product exponents and the maximum exponent.  A single EHU can be shared between multiple IPUs to amortize its overhead (i.e., multiplexed in time between IPUs), since a single FP-IP operation consists of multiple nibble iterations with the same exponent computation.

The range of the exponent for FP16 products is $[-28, 30]$, thus the exponent difference (i.e., the right shift amount) between two FP16 products can be up to 58-bit. In general, the bit width of the product increases based on the amount of right shift (i.e., alignment with the max exponent). However, due to the limited precision of the accumulator, an approximate computation is sufficient where the product alignment can be bounded and truncated to a smaller bit width. We define this width as \textbf{\emph{the IPU precision}} and use it to parametrize IPUs. The IPU precision is also the maximum amount of local right shift as well as the bit-width of the adder tree. We quantify the impact of precision on the computation accuracy in Section~\ref{sec:sw_accuracy}.

\textbf{The accumulator operations:}
During the computation for one pixel, FP accumulators keep two values: accumulator's exponent and its non-normalized signed magnitude. 
Once all the input vector pairs are computed and accumulated, the result in the accumulator is normalized and reformatted to the standard representation (i.e., FP16 or FP32).

The details of the accumulation logic are depicted in the right side of Figure~\ref{fig:IPU}. The accumulator has a $(33+t+l)$-bit register and a right shift unit (see Figure~\ref{fig:IPU} for defining $t$ and $l$). Therefore, the register size allows up to 33 bits of right shift, which is sufficient to preserve accuracy as discussed in Section~\ref{sec:sw_accuracy}.

In contrast to INT-mode accumulator, where the right shift logic can only shift by $4k$ ($k\in {1,2,..,6}$), the FP-IP can right shift by any number between [0:33+t+l]. The shift amount is computed in exponent logic and is equal to $4\times((3-i-1)+(3-j-1))+|max\_exp-exp|$, where $i$, and $j$ are input nibble indices, $exp$ is the accumulator's exponent value and $max\_exp$ is the adder tree exponent (i.e., the max exponent). A swap operation followed by a right shift is applied whenever a left shift is needed, hence, a separate left shift unit is not needed. In other words, the swap operation is triggered only when $max\_exp>exp$.

With respect to $exp$, the accumulator value is a fixed point number with $33+t+l$ bits, including sign, $(3 + t +l )$-bit in integer positions and 30 bits in fraction positions.  Note that the accumulator holds an approximate value since the least significant bits are discarded and its bit-width is provisioned for the practical size of IPUs. Before writing back the result to memory, the result is rounded to its standard format (i.e., FP16 or FP32).

For the rest of this paper, we define an $IPU(w)$ as an inner product unit with $5$-bit signed multipliers, $w$-bit adder tree, and local right shifter that can shift and truncate multipliers' output by up to $w$ bits.  We refer to $w$ as the IPU's adder tree precision or \textbf{IPU precision} for brevity.
In general, the result of $IPU(w)$ computation might be inaccurate, as only the $w$ most significant bits of the local shifter results are considered. However, there are exceptions:

  

\newtheorem{propos1}{Proposition}
\begin{propos1}\label{proposition:safe_precision}
  For $IPU(w)$, truncation is not needed and the adder tree result is accurate
  if the amount of alignments, given by EHU, of all the products are smaller than $w-9$. We refer to $w-9$ as \textbf{\emph{the safe precision of the IPU}}. 
\end{propos1}

It is clear that the area and power overhead increase as the IPU precision increases (See Section~\ref{sec:synthesis}). The maximum required precision is determined by the software requirement and the accumulator precision (See Section~\ref{sec:sw_accuracy}).

\section{Optimizing Floating Point Logic}
\label{sec:fp_op}

In this section, we tackle the overhead of large shifters and adder tree size by, first, evaluating the minimum shift and adder size required to preserve the accuracy (Section~\ref{sec:sw_accuracy}) for both FP16 and FP32 accumulators. Based on the evaluation, we propose optimization methods to implement FP IPUs with relatively smaller shift units and adders (Section~\ref{sec:mcipu} and Section~\ref{sec:ipu_cluster}).

\subsection{Precision Requirement for FP16}\label{sec:sw_accuracy}
\begin{figure}[!tp]
\vspace{-5pt}
\begin{center}
   \includegraphics[width=0.75\linewidth]{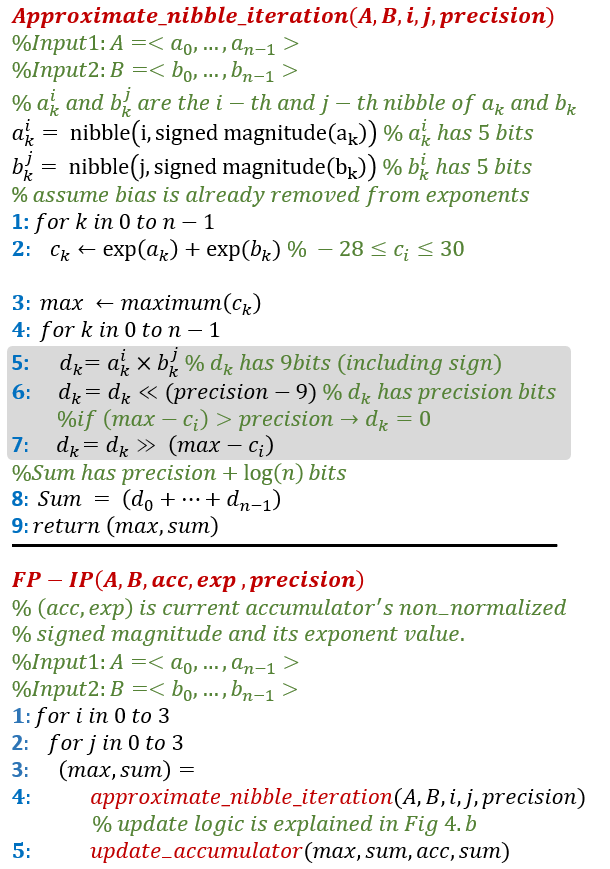}
\end{center}
\vspace{-7mm}
   \caption{
   Pseudocode for the approximate version of nibble iteration (top) and FP-IP operation with the approximate nibble iteration method (bottom).  Precision is the IPU precision.}
\vspace{-10pt}
\label{fig:FP-IPapprox}
\end{figure}

As we mention in Section~\ref{sec:ipu}, an FP-IP operation is decomposed into multiple nibble iterations. In a typical implementation, the multiplier's output of each iteration requires large alignment shifting and the adder tree has high precision inputs. However, this high precision would be discarded due to the limited precision of the accumulator (FP16 or FP32), hence, an approximated version of FP-IP alignment can be used without significant loss of  accuracy. 
Figure~\ref{fig:FP-IPapprox} shows the pseudocode for the approximate FP-IP operation customized for our nibble-based IPU architecture.
The approximate FP-IP computes only most significant $precision$ bits of the products (Lines 5-7). The $precision$ parameter allows us to quantify the absolute error. 

\newtheorem{theorm1}{Theorem}
\begin{theorm1}\label{theorem:abs_error}
  For FP-IP with $n$ pairs of FP16 inputs, the absolute error due to $approx\_nibble\_iteration(i,j,precision)$, called $abs\_error(i,j)$ is no larger than $225\times 2^{(4\times(i+j)-22)}\times2^{max-precision} \times (n-1)$, where $max$ is the maximum exponent of all the products in the FP operation.
\end{theorm1}

\textbf{Proof:} Due to space limitations, we only provide an outline of the proof. The highest error occurs when, except for one product, all $n-1$ others are shifted $precision$ to the right, and thus appear as errors. For maximum absolute error, these products should all have the same sign and have the maximum operand (i.e., 15). Hence their product would be $15\times 15=225$. The term $2^{(4\times(i+j))}$ is applied for proper alignment based on nibble significance. The term $2^{-22}$ is needed, since each FP number has 3-bit in int and 22-bit fraction positions, with respect to its own exponent.

\newtheorem{corol1}{Remark}
\begin{corol1}\label{Remark:abs_error}
  Iterations of the most significant nibbles (i.e., largest $i+j$) have the highest significant contributions to the absolute error.
\end{corol1}

\begin{figure*}[thp!]
\vspace{-9pt}
\begin{center}
   \includegraphics[width=\linewidth]{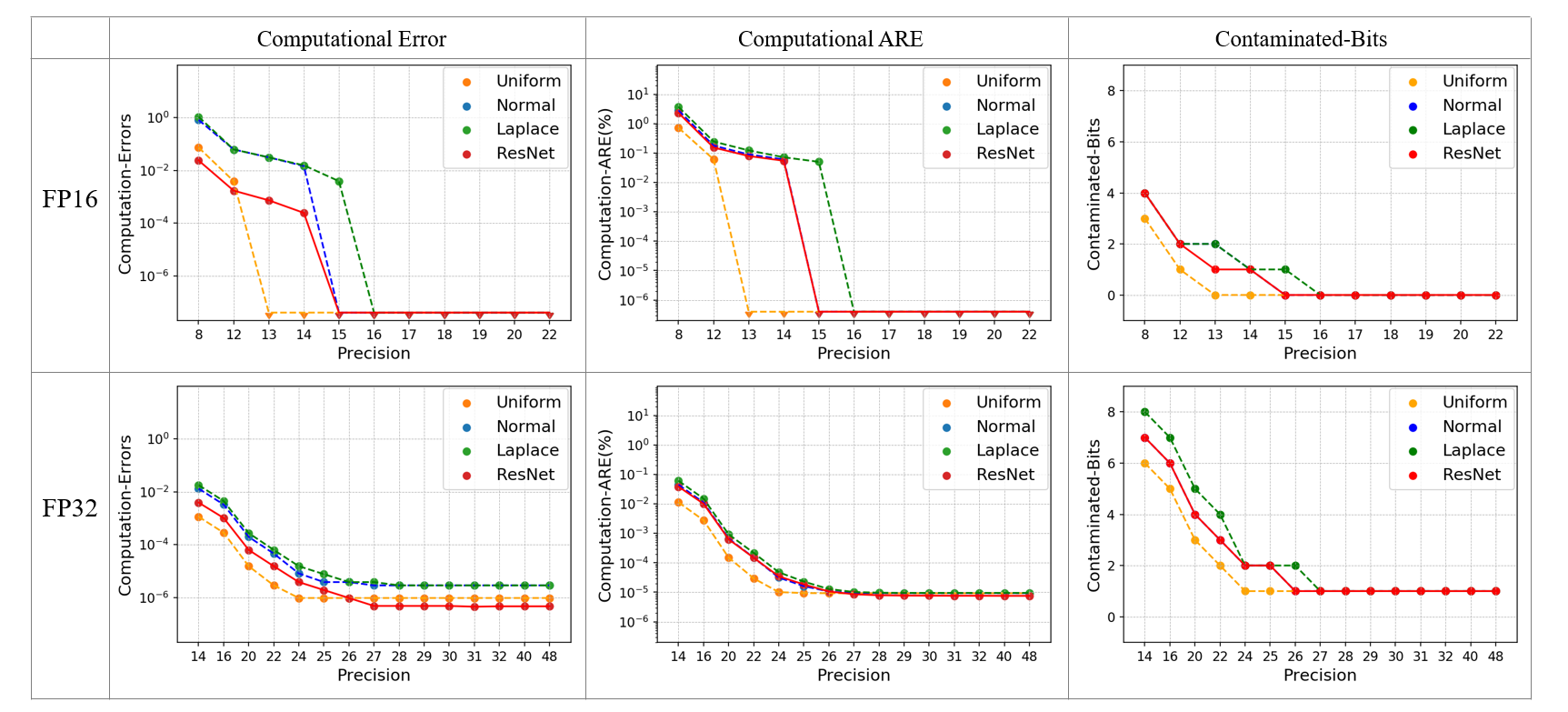}
\end{center}
\vspace{-20pt}
   \caption{
   Left to Right: Absolute error, percentage of absolute relative error (ARE), and the number of contaminated bits for different distributions and different accumulators: FP16(top) and FP32(bottom). The first two error graphs in each row use log-scale Y-axis. 
}
\vspace{-10pt}
\label{fig:sw_analysis}
\end{figure*}

The FP-IP operation is the result of nine approximate nibble iterations added into the accumulator. However, only 11 or 24 most significant bits of the accumulated result are needed for FP16 or FP32 accumulators, respectively. Unfortunately, the accumulator is non-normalized and its leading non-zero position depends on the input values. As a result, it is not possible to determine a certain precision for each approximate nibble iteration to guarantee any loss of significance. Therefore, we use numerical analysis to find the proper shift parameters. 
In our analysis, we consider both synthetic input values and input values sampled from tensors found in Resnet-18 and Resnet-50 convolution layers. We consider Laplace and Normal distributions to generate synthetic input vectors, as they resemble the distribution of DNN tensors~\cite{park2018value} and uniform distributions for the case that tensor is re-scaled, as suggested for FP16-based training~\cite{micikevicius2017mixed}. 
In our analysis, we consider 1M samples generated for our three distributions and 5\% data samples of Resnet-18 and Resnet-50 convolution layers.
For different IPU precisions, we measure the median for three metrics: absolute computation error, absolute relative error (in percentage) compared with FP32 CPU results, and the number of contaminated bits.  The number of contaminated bits refers to the number of different bits between the result of approximated FP-IP and the FP32 CPU computation.
Figure~\ref{fig:sw_analysis} include the error analysis plots for both FP16 and FP32 accumulator cases. 
Based on this analysis, we found that both the relative and absolute errors are less than $10^{-6}$ for 16-bit IPU precision in FP16 case. 
Moreover, the median number of contaminated bits is zero (mean = 0.5).
For accumulator in FP32  case, both errors drop to less than $10^{-5}$ for $IPU precision \ge 26$-bit. However, the minimum median value of the number of contaminated bits starts at 27b IPU precision.
We conclude that \textbf{ \emph{in order to maintain FP32 CPU accuracy, FP16 FP-IP operations require at least 16b and 27b IPU precision for accumulating into FP16 and FP32, respectively}}. 

We also evaluate the impact of IPU precision on Top-1 accuracy of ResNet-18 and ResNet-50 for ImageNet data set~\cite{He_2016_CVPR}. 
We observe that, when the FP16 uses IPU precision of 12 or more, it maintains the same accuracy (i.e., Top-1 and Top-5) as FP32 CPU for all batches.
IPU precision of 8-bit also shows no significant difference with respect to the final average accuracy compared to CPU computation.
However, we observe some accuracy drops of up to~17\% for some batches, and some accuracy improvements up to~17\% for other batches. 
We are not sure if this improvement is just a random behavior, or because lower precisions may have a regularization effect as suggested by~\cite{courbariaux2015binaryconnect}.
At any rate and despite these results, 8-bit IPU precision is not enough for all CNN inference due to the fluctuation in the accuracy for individual batches, compared to the FP32 model.


\begin{figure}
\vspace{-5pt}
\begin{center}
   \includegraphics[width=\linewidth]{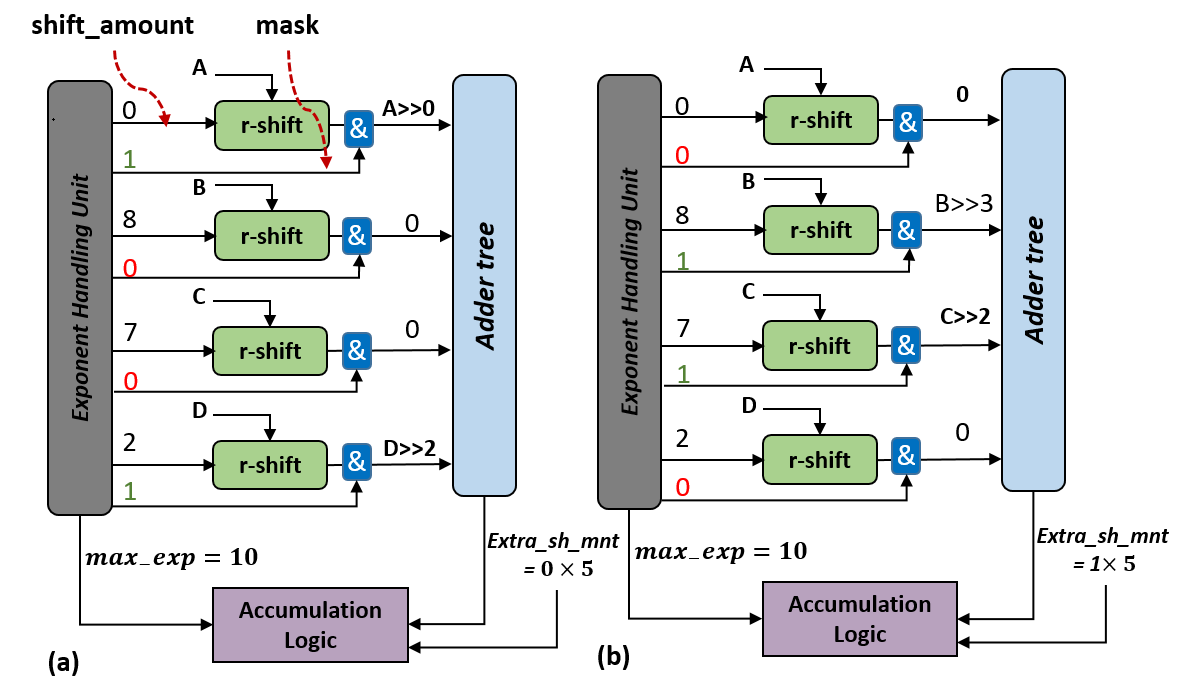}
\end{center}
    \vspace{-8mm}
   \caption{ Walk-through example for ($sp=5$) with (A,B,C,D) as magnitudes and (10,2,3,8) as exponents. The exponent can be written as (0,-8,-7,-2) with respect to $max\_exp=10$.
   (a) First cycle: MC-IPU only executes products A and D since their right shift is in $P_0=[0,5]$ (b) Second cycle: MC-IPU computes products B and C as their right shift is in $P_1=[5,10]$. 
}
\vspace{-15pt}
\label{fig:MCIPU}
\end{figure}

\subsection{Multi-Cycle IPU}
\label{sec:mcipu}

As we mentioned in Section~\ref{sec:sw_accuracy}, approximate nibble iteration requires 27-bit addition and alignment to maintain the same accuracy as CPU implementations for FP32 accumulation. 
As we illustrate in Section~\ref{sec:method}, the large shifter and adder take a big portion of area breakdown of an IPU and an overhead when running in the INT mode. 
In order to maintain both high accuracy and low area overhead, we propose using multiple cycles when a DNN requires large alignment, using \emph{multi-cycle IPU(w)}, (MC-IPU$(w)$),
where $w$ refers to the adder tree bit width.
Hence, designers can consider lower MC-IPU precision, in cases when the convolution tile is used more often in the INT than the FP mode.

MC-IPU relies on Proposition~\ref{proposition:safe_precision} that if all the alignments are smaller than the safe precision ($sp$), summation is accurate. Otherwise, the MC-IPU performs the following steps to maintain accurate computation. First,	it decomposes products into multiple partitions, such that products whose required shift amounts belong to $[k\times sp,(k+1)\times sp]$ are in partition $k$ ($P_k)$. Second, all products in partition $k$ are added in the same cycles and all other products are masked. Notice that all the products in $P_k$ require at least $k\times sp$ shifting. Thus MC-IPU decomposes the shift amount into parts: (1) $k\times sp$ that is applied after the adder tree and (2) the remaining parts that is applied locally. Since the remaining parts are all smaller than $sp$,  they can be done with local shift units without any loss in accuracy (Proposition~\ref{proposition:safe_precision}).

Figure~\ref{fig:MCIPU} illustrates a walk-through example for MC-IPU$(14)$, where $sp=5$.  
In this example, we denote the products in summation as A, B, C, and D with exponents 10, 2, 3, and 8, respectively. Thus, the maximum exponent is $max\_exp = 10$. Before the summation, each product should be aligned ($w.r.t. max\_exp$) by the right shift amount of 0, 8, 7, and 2, accordingly. The alignment and summation happens in two cycles as follows:
In the first cycle, A and D are added after zero- and two- bit right shifts, respectively. Notice that, the circuit has extra bitwise AND logic to mask out input B and C in this cycle.
In the second cycle, B and C are added and they need eight- and seven- bit right shifts, respectively. While the local shifter can only shift up to five bits accurately, we perform the right shift in two steps by locally shift by $(8-5)$ and $(7-5)$ bits, followed by five bit shifts of the adder tree result.

\begin{figure}[thp]
\vspace{-8pt}
\begin{center}
   \includegraphics[width=0.85\linewidth]{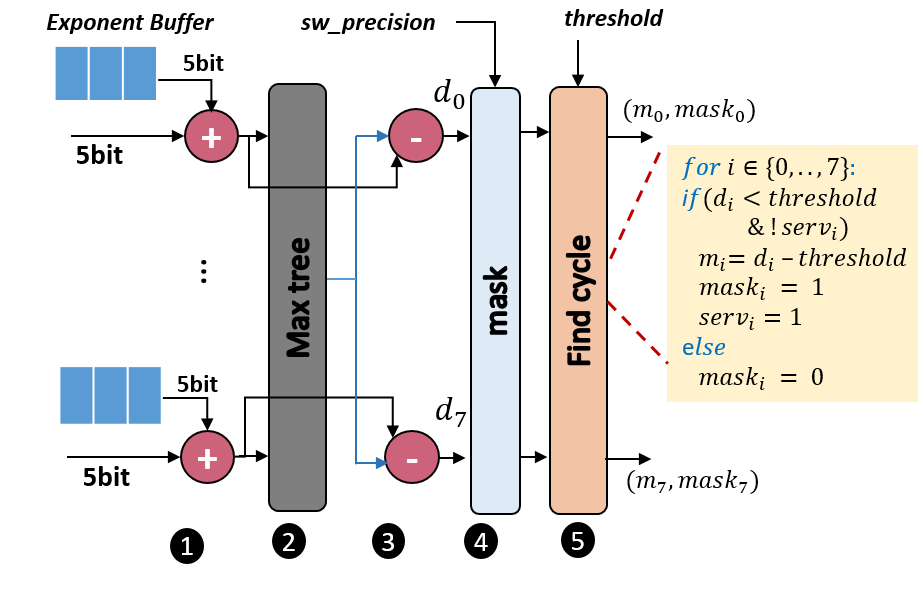}
\end{center}
\vspace{-25pt}
   \caption{EHU Data path for MC-IPUs.}
\vspace{-15pt}
\label{fig:EHU}
\end{figure}

In general, the Multi-Cycle IPU imposes three new overheads to IPUs: (1) Bitwise AND logic per multiplier; (2) updating shifting logic, where the shared shifting amount would be given to the accumulation logic (extra\_sh\_mnt in Figure~\ref{fig:MCIPU}, for each cycle; and (3) modifications to the EHU unit.
The EHU unit for MC-IPU is depicted in Figure~\ref{fig:EHU}.
It consists of five stages.
The first stage receives the activation exponent and weight exponents and adds them together to calculate the product exponents.
In the second and third stages, the maximum exponent and its differences from each product exponent are computed. 
In the fourth stage, the differences that exceed the software precision are masked (see Section~\ref{sec:sw_accuracy}).
The first four stages are common for both IPUs and MC-IPUs. 
However, the last stage is only needed for MC-IPU and might be called multiple times, depending on the required number of cycles for MC-IPU.
This stage keeps a single bit for each product to indicate whether that product has been aligned or not ($serv_i$ in Figure~\ref{fig:EHU}).
For the non-aligned ones, this stage checks the exponent difference value with a threshold.
The threshold value equals $(k+1)\times sp$ in cycle $k$ (see the code in Figure~\ref{fig:EHU}).
The EHU finishes for an FP-IP, once all products are aligned (i.e., $serv_i=1$). Notice that one EHU is shared between multiple MC-IPUs as it is need once for all nine nibble iterations.

\subsection{Intra-Tile IPU clustering}
\label{sec:ipu_cluster}

In the previous Section, we show how the MC-IPU can run the FP inner product by decomposing it into nibble iterations and computing each iteration in one or multiple cycles. In a convolution tile that leverages MC-IPUs, the number of cycles per iteration depends on two factors: (1) the precision of the MC-IPU (i.e., adder tree bit width). (2) the maximum alignment needed in all the MC-IPUs in the convolution tiles. When a  MC-IPU in the convolution tile requires a large alignment, it will stall others.


When architecting such an IPU, the first consideration is the INT and FP operations percentage split
The second factor, however, can be handled by grouping MC-IPUs in smaller clusters and running them independently. This way, if one MC-IPU requires multiple cycles, it stalls only the MC-IPUs in its own cluster. To run clusters independently, each cluster should have its own local input and output buffers. The output buffer is used to synchronize the result of different clusters before writing them back into the activation banks. Notice that the activation buffer broadcast inputs to each local input buffer and would stop broadcasting even if one of the buffers is full, which stalls the entire tile. 

\section{Methodology and Results}
\label{sec:method}

In this section, we illustrate the top level architecture and experiment setup. 
Then, We evaluate the hardware overhead and performance impact of our proposed architecture. We also discuss a comparison with some related work.

\subsection{Top Level Architecture}
\label{sec:top-level}
\begin{figure}
\vspace{-5pt}
\begin{center}
   \includegraphics[width=0.7\linewidth]{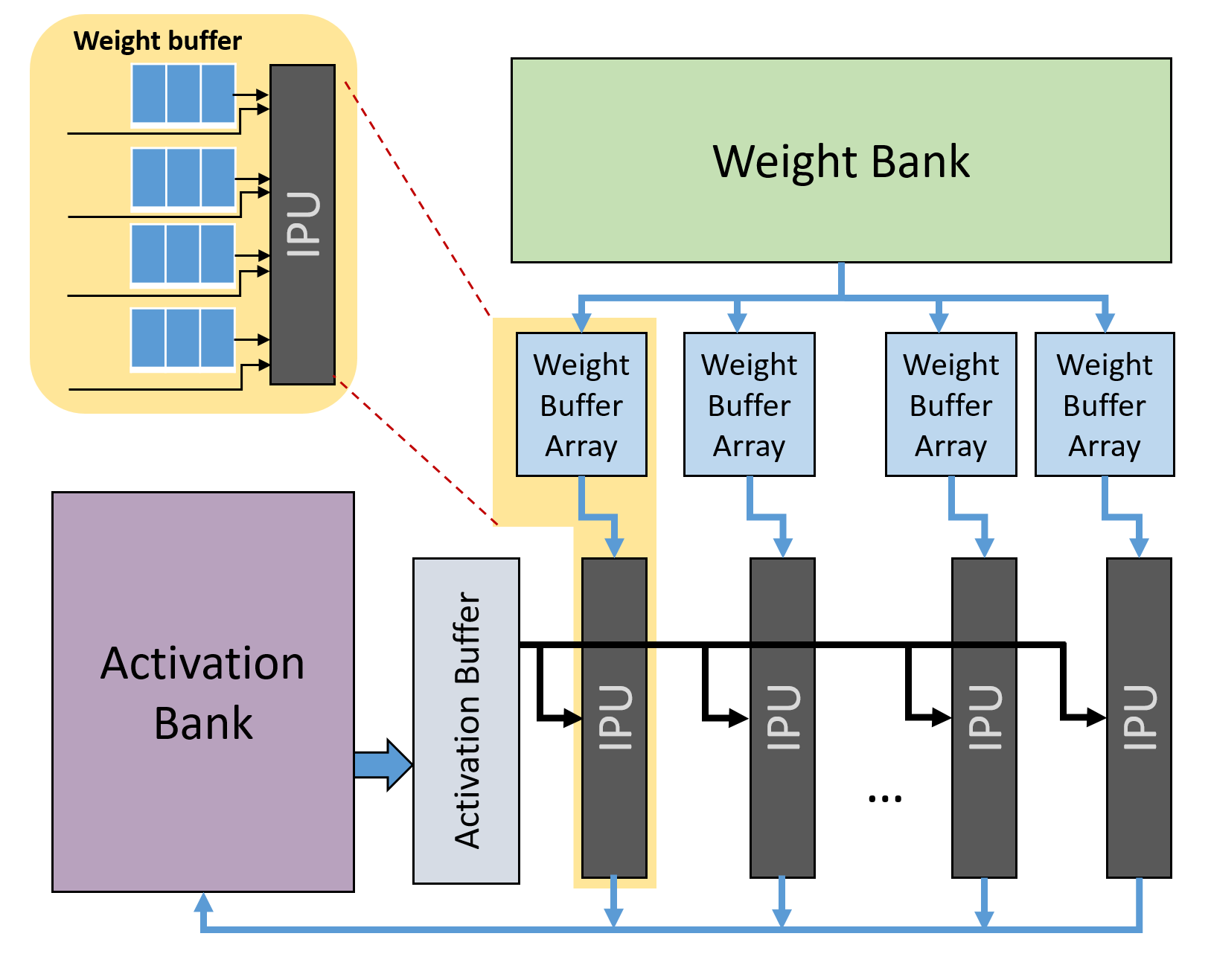}
\end{center}
    \vspace{-7mm}
   \caption{ Convolution tile architecture.} 
    \vspace{-20pt}
\label{fig:conv_tile}
\end{figure}

We consider a family of high-level architectures designed by IP-based tiles.
IP-based tiles are crucial for energy efficiency, especially when low-precision multipliers are used.
IP-based convolution tile consists of multiple IPUs and each IPU is assigned to one output feature map (OFM) (i.e., unrolling in output channel (K)).
All IPUs share the same input vectors that come from different channels in the input feature map (IFMs) (i.e., unrolling in the input channel dimension (C)).
As depicted in Figure~\ref{fig:conv_tile}(a), the data path of a convolution tile consists of the following components:
(1) Inner Product Unit: an array of multipliers that feeds into an adder tree. The adder tree's result is accumulated into the partial sum accumulator.
(2) Weight Bank: contains all the filters for the OFMs that are assigned to the tile.
(3) Weight buffer: contains a subset of filters that are used for the current OFMs. Each multiplier has a fixed number of weights, which is called the depth of the weight buffer. 
Weight buffer are only needed for weight stationary (WS)~\cite{chen2016eyeriss} architecture and is either implemented with flip-flops, register files, or small SRAMs. 
The number of elements per weight buffer determines the output/partial bandwidth requirements.
(4) Activation Bank: contains the current activation inputs, partial, and output tensors.
(5) Activation Buffer: serves as a software cache for the activation bank.

We consider, two types of tiles, big and small, based on INT4 multipliers. Both tiles are weight stationary with weight buffer depth of 9B. The big and small tiles are unrolled $(16,16,2,2)$  and $(8,8,2,2)$ in $(C,K,H_,W_o)$ dimensions. 
We consider these two tiles because they offer different characteristics while achieving high utilization. 
The IPUs in the big tile have twice as many multipliers as in the small tile (16 vs. 8). The 16-input IPUs have smaller accumulator overhead but larger likelihood of multiple cycles alignment as compared to 8-input IPUs.
For comparison, we consider two baselines: Baseline1 and Baseline2 for the small and the big tiles, respectively.
Each baseline has four tiles with a 38b wide adder tree per IPU. 
Hence, these baselines do not need MC-IPU (Section~\ref{sec:mcipu}) and IPU clustering (Sectoin~\ref{sec:ipu_cluster}) and they can achieve (1~TOPS, 113~GFLOPS) and (4~TOPS, 455~GFLOPS), respectively (OP is a $4\times4$ MAC). 

The performance impact of the proposed designs (i.e., MC-IPUs and clustering the IPUs) depends on the distribution of inputs.
We developed a cycle-accurate simulator that models the number of cycles for each convolution layer. 
The simulation parameters include the input and weight tensors.
The simulator receives, the number of tiles, the tile dimension (e.g., (8, 8, 2, 2) for the small tiles), and the number of clusters per tile.
We simulate Convolution layers as our tiles are customized to accelerate them.
In addition, we assume an ideal behavior for the memory hierarchy to single out the impact of our designs.
In reality, non-CNN layers and system-level overhead can impact the overall result.
Moreover, the area and power efficiency improvements might decline due to the limitations of DRAM bandwidth and SRAM capacity~\cite{Dark_Mem}.
Such scenarios are beyond the scope of our analysis.

In the simulation analysis, we use data tensors from ResNet~\cite{He_2016_CVPR} and InceptionV3~\cite{Szegedy_2016_CVPR}.
We consider four study cases which are: (1) ResNet-18 forward path, (2) ResNet50 forward path, (3) InceptionV3 forwad path, and (4) ResNet-18 backward path of training.
In our benchmarks, we consider at least 16b and 28b software precision (Section~\ref{sec:sw_accuracy}) that is required for FP16 and FP32 accumulation to incur no accuracy loss.

\begin{figure*}[t]
\vspace{-8pt}
\begin{center}
   \includegraphics[width=.85\linewidth]{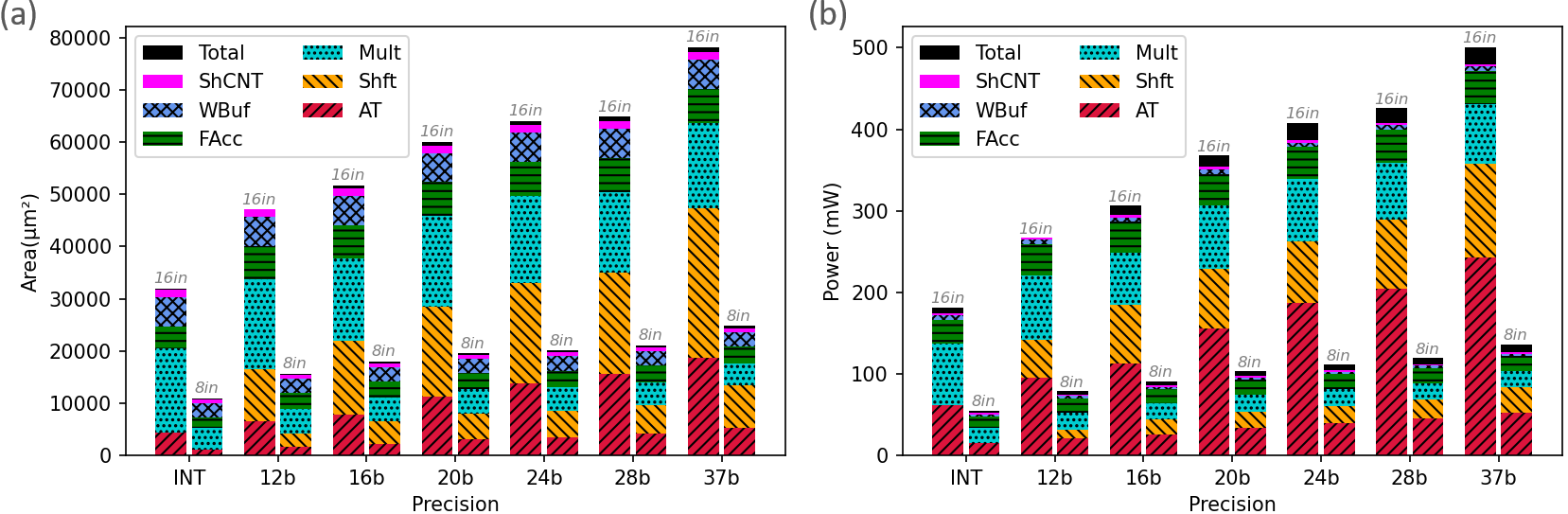}
\end{center}
    \vspace{-7mm}
   \caption{
   Breakdown of (a) area (b) power for different MC-IPU based tiles. The components are accumulators (FAcc), weight buffers (WBuf), EHUs (ShCNT), multipliers (MULT), local shifters (Shft), and adder trees (AT).
}
\vspace{-3mm}
\label{fig:synthesis}
\end{figure*}

\subsection{Hardware Implementation Results}
\label{sec:synthesis}
In order to evaluate the impact of FP overheads, we implemented our designs in SystemVerilog and synthesized them using Synopsys DesignCompiler with $7nm$ technology libraries~\cite{DC}. We consider 25\% margin and 0.71$V$ Voltage for our synthesis processes. 
Figure~\ref{fig:synthesis} illustrates the breakdown of area and power for a small and big tile. We also include a design point without FP support, shown as INT in Figure~\ref{fig:synthesis}. In addition, we consider one design with a 38-bit adder tree, similar to NVDLA~\cite{NVDL}, for our baseline configuration. 
We highlight the following points in Figure~\ref{fig:synthesis} as follows:
(1) By just dropping the adder tree precision from 38 to 28 bits, which is the minimum precision to maintain CPU-level accuracy for FP32 accumulations (see Section~\ref{sec:sw_accuracy}), the area and power are reduced by 17\% and 15\% for 16-input and 8-input MC-IPU tiles, respectively.  
(2) We can reduce the adder tree precision even further at the cost of running alignment in multiple cycles. The tile area can be reduced by up to 39\% when reducing adder tree precision to 12 bits.
(3) In comparison with INT only IPU, MC-IPU(12) can support FP16 with a $43\%$ increase in area.



\begin{figure}
\vspace{-10pt}
\begin{center}
   \includegraphics[width=.8\linewidth]{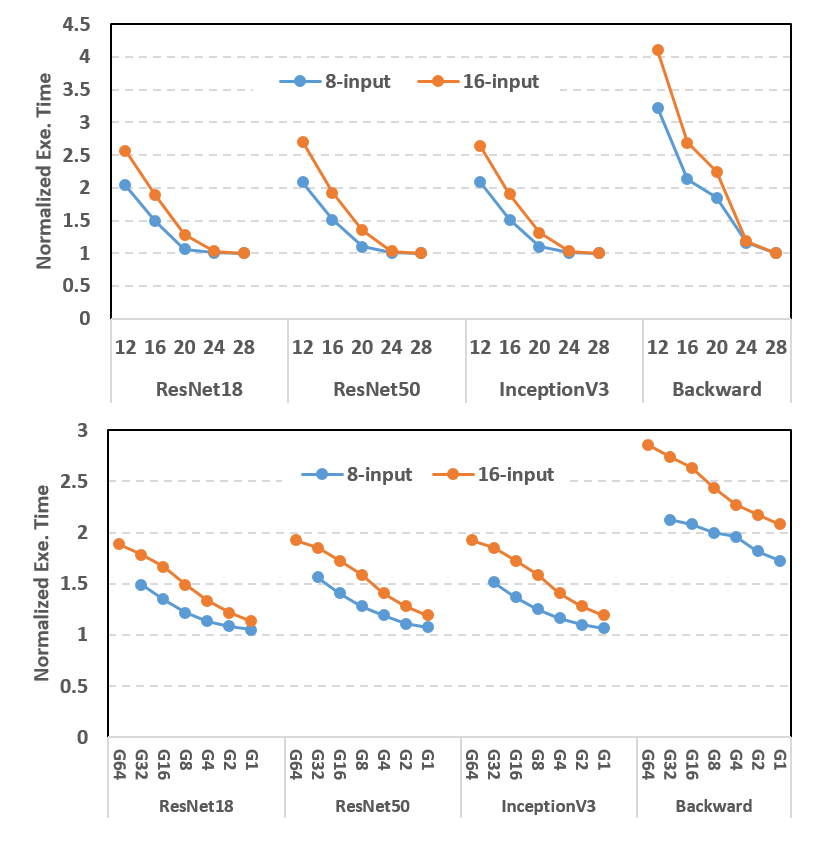}
\end{center}
   \vspace{-25pt}
   \caption{
   (a) Impact of different precision on the performance of MC-IPUs. Backward refers to the back propagation error in ResNet-18.(b) Impact of cluster size on the performance for MC-IPU(16).
}
   \vspace{-20pt}
\label{fig:mcpi-perf}
\end{figure}

\subsection{Performance Result}
\label{sec:performance}
\textbf{FP16 operations with FP16 accumulations}: As shown in Section~\ref{sec:sw_accuracy}, there is no need for more than 16-bit precision for FP16 accumulation. 
Therefore, IPUs with a 16b or larger adder tree take exactly one cycle per nibble iteration. However, MC-IPU(12) may require multiple-cycle alignment execution, which causes  performance loss. Compared to Baseline1 (Baseline2), when MC-IPU(12)s are used, the performance drops by 47\% (50\%), on average,  when no IPU clustering is applied (Section~\ref{sec:ipu_cluster}). 
If we choose a cluster of size one, (i.e., MC-IPUs perform independently),  the performance drop is 26\% (38\%), compared to Baseline1 (Baseline2).


\textbf{FP16 operations with FP32 accumulations}: 
As we mentioned in Section~\ref{sec:sw_accuracy}, FP32 accumulation requires 28-bit IPU precision. Thus, an MC-IPU with precision less than 28-bit might require multiple cycles, causing performance loss.
 Figure~\ref{fig:mcpi-perf} shows the normalized execution time for different precision values for the forward path of ResNet-18, ResNet-50, and InceptionV3 as well as the backward path of ResNet-18.
 We observe that all epochs have almost similar trend, thus we only report data for Epoch 11. 
 In this figure, we present two sets of numbers: ones for the tiles with 8-input MC-IPUs, normalized to Baseline1 and one for the tiles with 16-input MC-IPUs, normalized to Baseline2.

According to Figure~\ref{fig:mcpi-perf}~(a), the execution time can increase dramatically when small adder trees are used and 28-bit IPU precision is required.
The increase in the latency can be more than $4\times$ for a 12b adder tree in the case of computation of back propagation~(backprop).
Intuitively, increasing the adder bit width reduces the execution time. In addition, since 8-input MC-IPUs have fewer products, it is less likely that they need multiple cycles.
Thus, 8-input MC-IPUs~(Baseline1) outperform 16-input MC-IPUs~(Baseline2). We also observe that backprop computations have more dynamic range and more variance in the exponents.

\begin{figure}[t]
\vspace{-8pt}
\begin{center}
   \includegraphics[width=\linewidth]{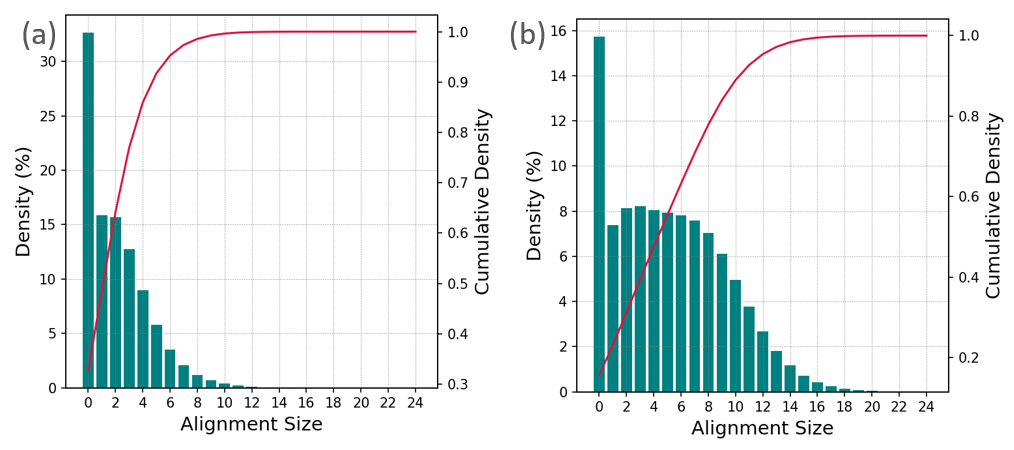}
\end{center}
  \vspace{-20pt}
   \caption{
   The distribution of exponent difference ($Max.exp-exp$, or alignment size) of ResNet-18 training computations. (a) forward-propagation, (b) back-propagation.
}
\vspace{-10pt}
\label{fig:expdifference}
\end{figure}

To evaluate the effect of clustering, We fix the adder tree bit-width to 16 bits and vary the number of MC-IPUs per cluster.
Figure~\ref{fig:mcpi-perf}~(b) shows the efficiency of MC-IPU clustering, where the x-axis and y-axis represents the cluster size and the execution of 8-input (16-input) MC-IPUs(16) normalized to Baseline1 (Baseline2) respectively.
According to this figure, smaller clusters can reduce the performance degradation significantly due to multi-cycling in the case of forward computation using 8-input MC-IPUs. However, in 16-input cases, there is at least 12\% loss even for cluster of size 1. 
Backward data has more variation and, even for one MC-IPU per cluster, there is at least 60\% increase in the execution time.
The reason for such behavior can be explained using the histogram of exponent difference of 8-input MC-IPUs for Resnet-18 in the forward and backward paths, illustrated in Figure~\ref{fig:expdifference}.
As shown in this figure, the forward path exponent differences are clustered around zero and only 1\% of them are larger than eight. On the other hand, the products of backward computations have a wider distribution than forward computations.

\begin{figure}[t]
\vspace{-2pt}
\begin{center}
   \includegraphics[width=0.8\linewidth]{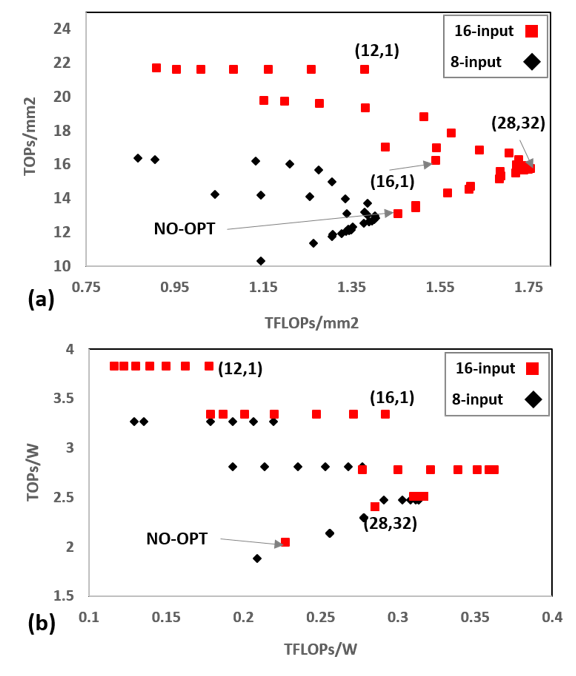}
\end{center}
\vspace{-25pt}
   \caption{
   Trade-off between (a) area efficiency and (b) power efficiency. Each design point (\emph{p},\emph{c}) represents tiles with the \emph{p}-bit adder tree MC-IPUs with \emph{c} MC-IPUs per cluster (only labeled for 16-input MC-IPUs). NO-OPT is Baseline2.
}
\vspace{-20pt}
\label{fig:trade-off}
\end{figure}


\subsection{Overall Design Trade-offs} 
\label{sec:tradeoffs}
Figure~\ref{fig:trade-off}(a,b)~visualize the power and area efficiency design spaces for INT vs. FP modes, respectively.
In these figures, we consider the average effective throughput, using our simulation results, for FP throughput values. 
The numbers associated with some of the design points refer to the ordered pair of MC-IPU precision and the cluster size. 
For designs with 8-input (16-input), approximation can boost power efficiency of INT and FP mode by 14\% (17\%) and improve area efficiency by 17.8\% (20\%). The overall improvement is the combination of all the optimizations. 
The two design points (12,1) and (16,1) are on the power efficiency Pareto optimal curve.
For example, the design points with one MC-IPU per cluster and 12-bit (16-bit) IPU precision, achieve 14\% (25\%) in TFLOPS/$mm^2$ and up to 46\% (46\%) in TOPS/$mm^2$ with our 8-input (16-input) IPU architectures over typical mixed precision implementation in area efficiency and up to 63\% (40\%) in TFLOPS/W and up to 74\% (63\%) in TOPS/W in power efficiency.

\subsection{Sensitivity Analysis}
\label{sec:Sensitivity}
In this paper, we mainly consider INT4 as the common case, however, it is still possible to consider different precision as the baseline for different targeted quantization schemes, data types, application domain (i.e., edge vs cloud) and DNNs.
Therefore, we evaluate the performance of the proposed approach using four designs with different multiplier precisions. 
The first design (MC-SER) is based on serial multipliers (i.e., $12\times1$) similar to Stripes~\cite{judd2016stripes} but MC-SER supports FP16 using the proposed optimizations.
Note that, FP16 operation requires at least 12 cycles per inner product in the case of $12\times1$ multiplier. 
The second design (MC-IPU4) is optimized for INT4 as discussed earlier and it is based on $4\times4$ multipliers.
The third design (MC-IPU84) is optimized for INT8 for activation and INT4 for weights, and it is based on  $8\times4$ multipliers. 
The fourth design (MC-IPU8) is optimized for INT8 for activation and weights, and it is based on  $8\times8$ multipliers. 
We also compare against other mixed precision designs including: NVDLA, typical FP16 implementation and mixed precision INT-based designs which do not support FP16.
We show the comparison between these designs in terms of TOPS/mm$^2$ and TOPS/W for different types of operations as shown in Table~\ref{tb:mult}.
The results show that MC-IPU mitigates the overhead of the local shift units and adder trees when FP16 is required. This overhead becomes relatively more significant as the precision of the multiplier decreases and the optimization benefit decreases as we increase the baseline multiplier precision. 
However, designs with high multiplier baseline (e.g., $8\times 8$) limits the benefits of low-bit (e.g., INT4) software optimization.

\begin{table*}

\vspace{-15pt}
\centering
\scriptsize
\caption{TOPs/W and TOPs/mm$^2$ for different multipliers (MUL) and adder trees (ADT) precision. A and W are activation and weight precisions}
\vspace{0pt}

\begin{tabular}{ |c|c|c|c|c|c|c|c|c| } 
 \hline
  & MC-SER & MC-IPU4 & MC-IPU84 & MC-IPU8  & NDVLA &  FP16 & INT8 & INT4 \\ 
  
 \hline\hline
 ADT & 16b & 16b & 20b & 23b  & 36b &  36b & 16b & 9b \\
 \hline
 MUL & $12\times1$ & $4\times4$ & $8\times4$ & $8\times8$ &$8\times8$  &$12\times12$  &$8\times8$ &$4\times4$ \\
 \hline
 $A \times W$ & \multicolumn{8}{c|}{$TOPS/mm^2$ or $TFLOPS/mm^2$}\\
 \hline \hline
 $4\times4$         & 5.5 & 18.8 & 14.3 & 11.4 & 9.7 & 6.9 & 18.5 & 30.6  \\ 
 $8\times4$         & 5.5 & 9.4  & 14.3 & 11.4 & 9.7 & 6.9 & 18.5 & 15.3  \\ 
 $8\times8$         & 2.8 & 4.7  & 7.2  & 11.4 & 9.7 & 6.9 & 18.5 & 7.7   \\ 
 $FP16\times FP16$  & 0.9 & 1.6  & 1.8  & 5.4  & 4.9 & 6.9 &  --  & --  \\ 
 \hline \hline
 $A \times W$ & \multicolumn{8}{c|}{$TOPS/W$ or  $TFLOPS/W$}\\
 \hline \hline
 $4\times4$         & 1.4  & 3.3 & 2.4  & 1.8  & 1.5  & 0.9 & 2.8  & 5.6 \\ 
 $8\times4$         & 1.4  & 1.7 & 2.4  & 1.8  & 1.5  & 0.9 & 2.8  & 2.8\\ 
 $8\times8$         & 0.7  & 0.8 & 1.2  & 1.8  & 1.5  & 0.9 & 2.8  & 1.4\\ 
 $FP16\times FP16$  & 0.2  & 0.3 & 0.3  & 0.8  & 0.7  & 0.9 & --   & -- \\ 
 \hline
\end{tabular}    
\vspace{-15pt}
\label{tb:mult}
\end{table*}
\section{Related Work}
\label{sec:related}
Previous studies on CNN accelerators exploit two major approaches to their ALU/FPU datapath, MAC-based~\cite{jouppi2017datacenter,chen2016eyeriss,gao2017tetris,lu2017flexflow,kim16,Venkataramani17,Yazdanbakhsh18} and Inner Product-based~\cite{chen2014dadiannao,NVDL,Ethos,Venkatesan19,shao2019simba,liu2016cambricon,kwon2018maeri}.
Unfortunately, most of these approaches exploit INT-based arithmetic units and rely on quantization to convert DNNs from FP to INT.




The INT-based arithmetic unit can also support different bit widths. Multi-precisions of operands for INT-based architectures has been already addressed in both spatial and temporal decomposition. 
In the spatial decomposition approach, a large arithmetic unit is decomposed into multiple finer grain units~\cite{sharma18,Camus19,mei19,moons17}.
Since the Pascal architecture, Nvidia GPUs implement spatial decomposition via DP4A and DP2A instructions, where INT32 units are decomposed into 4-input INT8 or 2-input INT16 inner products.
This approach is different than ours, as we support FP16 and use inner product rather than MAC units.
On the other hand, the temporal decomposition approach performs the sequences of fine-grain operations in time to mimic a coarse-grain operation.
Our approach resembles this approach with 4-bit operations as the finest granularity.
Other works that use this approach prefer lower precision~\cite{judd2016stripes,lee19,eckert2018neural,sharify2018loom}. 
Temporal decomposition has also been used to avoid ineffectual operations by dynamically detecting fine-grain zero operands and discarding the operation~\cite{delmas2018bit,Albericio17,sharify2019laconic}. In contrast to us, these approaches do not support FP16 operands. 
In addition, we only discuss the dense architectures; however, the fine-grain building block can also be used for sparse approaches. We leave this for future. 
The approaches listed above rely on quantization schemes to convert FP32 DNNs to integer-based ones~\cite{krishnamoorthi2018quantizing,lee2018quantization,nagel2019data,zhuang2018towards,wang2018two,choi2018pact,hubara2017quantized}. These schemes are added to DNN software frameworks such as TensorFlow Lite. Recent advancements show that 8-bit post-training quantization~\cite{jacob2018quantization} and 4-bit retaining-based quantization can achieve almost the same performance as FP32~\cite{jung2019learning}. However, achieving high accuracy is less trivial for shallow networks with 2D Convolution operations~\cite{howard17,sheng18}. There is also work to achieve high accuracy at lower precision~\cite{zhu16,Zhuang19,banner19,Choukroun19,Courbariaux15,Zhou16, zhang2018lq,rastegari2016xnor}. A systematic approach to find the correct precision for each layer has been shown in~\cite{Wang_2019_CVPR,Dong_2019_ICCV,cai2020zeroq}. Dynamic multi-granularity for tensors is also considered as a way of computation saving~\cite{shen2020fractional}. Several quantization schemes have been proposed for training~\cite{wu2018training,banner2018scalable,das2018mixed,de2018high,park2018value}. 



Recent industrial products support mixed-precision arithmetic, including Intel's Spring Hill~\cite{springhill}, Huawei’s DaVinci~\cite{davinci}, Nvidia’s TensorCore~\cite{tensorcore}, Google's TPU~\cite{jouppi2017datacenter}, and Nvidia’s NVDLA~\cite{NVDL}. While most of these architectures use FP16, BFloat16 and TF32 are selected for the large range in some products~\cite{abadi2016tensorflow,tf32}. 
Using the current structure, our approach can support both BFloat16 and TF32 by modifying the EHU to support 8-bit exponents and using only four nibble iterations. Similar to our approach, NVDLA supportd FP-IP operations. 
In contrast, it decomposes an FP16 unit into two INT8 unit spatially.
Additionally, NVDLA does not allow computations of FP-IP operations with different-type operands.
It also does not support INT4 and provides a large precision (38-bit) for its adder tree, which we demonstrate it is not efficient.
Our proposed architecture optimization can also applied to spatial decomposition and it is orthogonal to the decomposition scheme (i.e., temporal, serial, spatial). 

There are also proposals to optimize the microarchitecture of FP MACs or IPUs. LMA is a modified FP units that leverages Kulisch accumulation to improve FMA energy efficiency~\cite{johnson2018rethinking}. An FMA unit with fixed point accumulation and lazy rounding is proposed in~\cite{brunie17}.
A 4-input inner product for FP32 is proposed in~\cite{Sohn16}. The spatial fusion for FMA is presented in~\cite{Zhang19}. Finally, a mixed precision FMA that supports INT MAC operations is presented in~\cite{zhang20}. As opposed to the proposed architecture, most of these efforts do not support INT-based operations or are optimized for FP operation with high overhead that hinder the performance of the INT operations.

\section{Conclusion}
\label{sec:conclusion}


In this paper, we explored the design space of the structure of an inner product based convolution tile and identified the challenges to support the floating-point computation and its overhead.
Further, from the software perspective, we investigated the minimum requirements for achieving the targeted accuracy.
We proposed novel architectural optimizations that mitigate the floating-point logic overheads in favor of boosting computation per area for INT-based operations.
We showed that for an IPU based on low-precision multipliers, adder and alignment logic overhead due to supporting FP operations is substantial. We conclude that the differences between product exponents are typically smaller than eight bits allowing the use of smaller shift units in FPUs.

\newpage

\bibliography{refs}
\bibliographystyle{mlsys2020}

\clearpage
\appendix
\section{Background}\label{sec:background}

\subsection{Convolution Layer Operation}\label{sec:conv_op}
\begin{figure}[h]
\begin{center}
   \includegraphics[width=\linewidth]{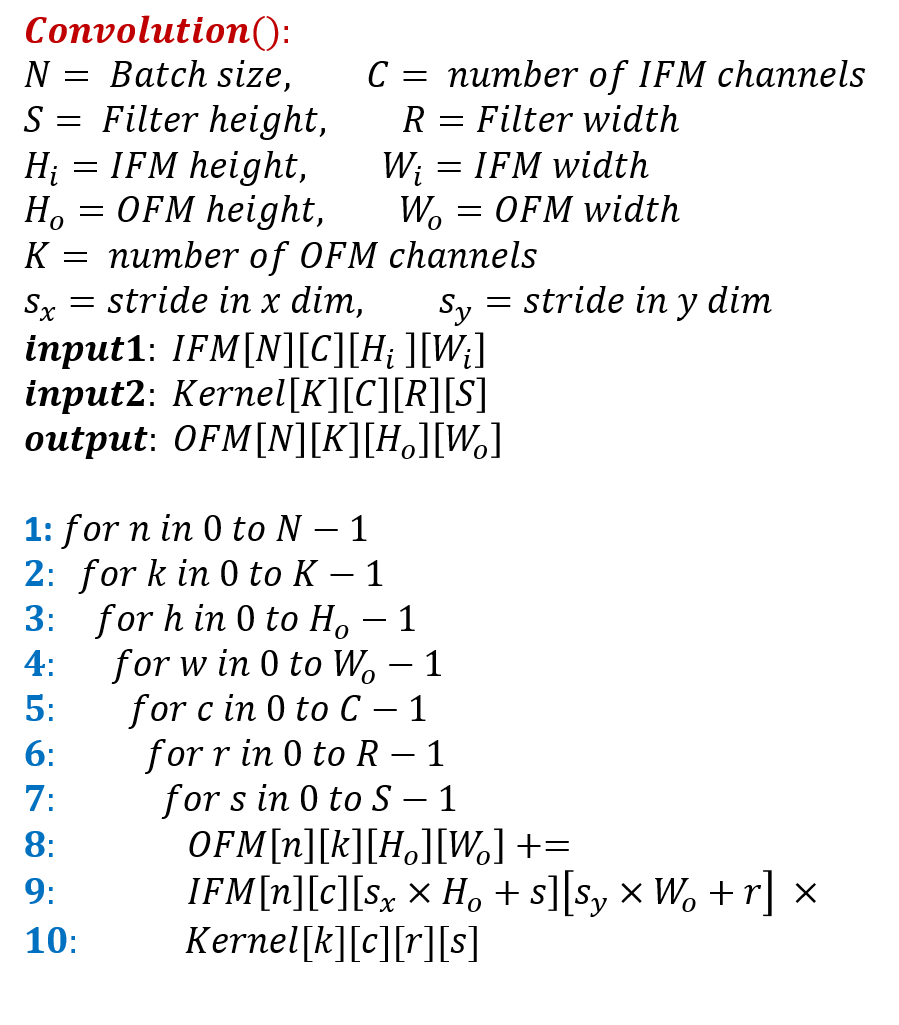}
\end{center}
   \caption{
   Pseudocode of a convolution layer. 
}
\label{fig:conv_op}
\end{figure}

A typical Convolution Layers (CL) operates on two 4D tensor as inputs (Input Feature Map (IFM) tensor and Kernel tensor) and results a 4D tensor (Output Feature Map (OFM) tensor). The element of IFMs and OFMs are called \emph{pixels} or \emph{activations} while the elements of Kernel are known as \emph{weights}. Figure~\ref{fig:conv_op} shows simplified pseudocode for CL.
The height and width of an OFM is typically determined by the height and width of IFMs, padding and strides. The three innermost loops (Lines 5-7) compute one output pixel and they can be realized as one or multiple inner product operations. The other four loops are independent, hence they can be implemented so to boost parallelism.
More details are presented in \cite{dumoulin2016guide}.
A fully connected layer can be considered as a special case of convolution where the height and the width of IFM, OFM and Kernel are all equal to 1. Fully connected layers are used frequently in natural language processing and in the final layers of Convolutional Neural Networks (CNNs). 

\subsection{Floating-Point Representation}
\label{sec:fp_basics}
Typical floating-point (FP) numbers are an IEEE standard to represent real numbers~\cite{8766229}. DNNs take advantage of floating point arithmetic for training and highly accurate inference tasks. In general, an FP number is represented with three parts: (sign, exponent, and mantissa), which have (1,5,10), (1,8,23), (1,8,7), and (1,8,10) for FP16, FP32, Google's BFloat (BFloat16)~\cite{abadi2016tensorflow} and Nvidia's TensorFloat32 (TF32)~\cite{tf32}.


For IEEE standard FP, the (sign, exponent, and mantissa) parts are used to decode five types of FP numbers as shown in Table~\ref{tb:fp}. We define the \textbf{\emph{magnitude}} as 0.mantissa for subnormal numbers and 1.mantissa for normal numbers. We also call it the \textbf{\emph{signed magnitude}} when signed values are considered.

\begin{table}

\caption{Different types of FP numbers. ($exp\neq0$, $exp\neq1...1$). $bias=15(127)$ for FP16 (FP32).}
\footnotesize
\begin{tabular}{ |c|c|c| } 
 \hline
 type & {$(sgn, exp, man)$} & Value \\ 
 \hline\hline
 zero & $(sgn, 0...0, 0...0)$ & zero \\ 
 INF & $(sgn, 1...1, 0...0)$ & $\pm$ infinity \\ 
 NaN & $(sgn, 1...1, man)$ & $man\neq0$, Not-a-Number  \\ 
 normal & $(sgn, exp, man)$ & $(-1)^s\times2^{exp-bias}\times1.man$ \\ 
 subnormal & $(sgn, {0...0} man)$ & $(-1)^s\times2^{-bias+1}\times0.man$ \\
 \hline
\end{tabular}
\label{tb:fp}
\end{table}

\begin{figure}[htbp]
\vspace{0pt}
\begin{center}
   \includegraphics[width=\linewidth]{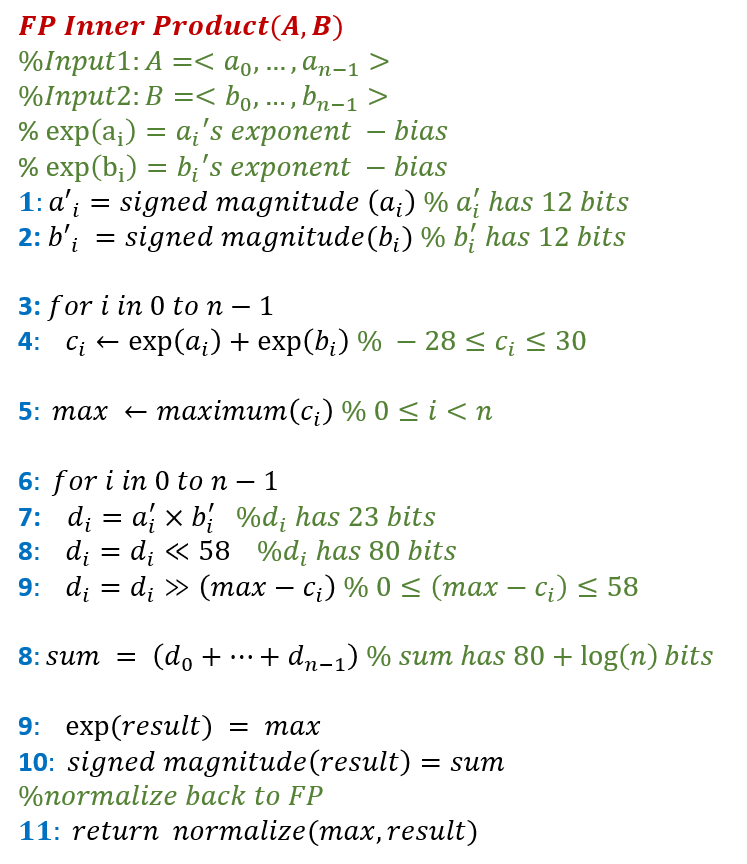}
\end{center}
    \vspace{-15pt}
   \caption{
    Pseudocode for FP-IP operation (FP16). In a hardware realization, the loops would be parallel. Note, $exp(x) = x's exponent -bias +1$ for subnormal numbers but we omit it for simplicity.
    }
\label{fig:FP-IP}
\end{figure}

For deep learning applications, the inner product operations can be realized in two ways: (1) by iteratively using fused-multiply-add (FMA) units, i.e., performing $A\times B+C$ or (2) by running multiple inner product operations in parallel. In the latter case, the inputs would be two vectors $\langle a_0,…,a_{n-1} \rangle$ and $\langle b_0,…,b_{n-1}\rangle$  and the operation results in one scalar output. In order to keep the most significant part of the result and guarantee an absolute bound on the computation error, the products are summed 
by aligning all the products relative to the product with the maximum exponent. Figure~\ref{fig:FP-IP} shows the required steps, assuming there is neither INF nor NaN in the inputs. The result has two parts: an exponent which is equal to the maximum exponent of the products, and a signed magnitude part which is the result of the summation of the aligned products.

The range of the exponent for FP16 numbers is [-14,15], hence, the range of the exponent for the product of two FP16 number is [-28,30]. The product result also has up to 22 bits of mantissa before normalization. This means that the accurate summation of such numbers requires 80-bit wide adders (58+22=80). However, smaller adders might be enough depending on the accuracy of the accumulators. For example, FP32 accumulators may keep only 24 bits of the result's sign magnitude. Therefore, it is highly unlikely that the least significant bits in the 80-bit addition contribute to the 24 bit magnitude of the accumulator and an approximate version of this operation would be sufficient. We will discuss the level of approximation in Section~\ref{sec:sw_accuracy}.

\section{Hybrid DNNs and Customized FP}

The temporal INT4-based decomposition allows the proposed architecture to support different data types and precisions per operand per DNNs’ layer. In the case that at least one of the operands is FP, the IPU runs in the FP mode. Depending on the input data types, the convolution results would be accumulated in a large INT or non-normalized FP register, which should be converted back to the next layer precision (INT or FP16 type). The conversion unit is not part of the IPU and thus not in the scope of this paper.

The proposed architecture can also support custom FP format, as we mentioned in Section~\ref{sec:fp_basics}, BFloat16 and TF32 have 8-bit exponents. We can support these types with two modifications. (i) The EHU should support 8-bit exponents and (ii) larger shift units and adders might be needed. 

Beside FP16 and BFloat16, there are some efforts to find the most concise data type for DNN applications. 
Flexpoint is a data type at the tensor level, where the all the tensor elements share an exponent and are 2s complement numbers~\cite{koster2017flexpoint}. The same concept is used in~\cite{drumond2018training,cambier2020shifted}.
Some studies shows how to train using shared exponent and FP. Deft-16 is introduced to reduce memory bandwidth for FP32 training~\cite{hill17}. Posit introduces a new field, called regime, to increase the range of numbers~\cite{gustafson2017beating, lu2019training}. 
Other studies show how to train using shared exponent and FP. Deft-16 is introduced to reduce memory bandwidth for FP32 training~\cite{hill17}. Posit introduced a new field, called regime, to increase the range of numbers and shows efficacy in DNN training as well ~\cite{gustafson2017beating, lu2019training}. 
Custom floating point representations are also proposed and they can be more effective compared to INT quantization in compressing DNNs with wide weight distributions such as transformers~\cite{tambe2019adaptivfloat}.


\end{document}